\documentclass[showpacs,amsmath,reprint,showkeys,pra,longbibliography]{revtex4-2}
\usepackage{dcolumn}    
\usepackage{multirow}
\usepackage{bm}         
\usepackage{ifpdf}
\usepackage{amssymb,lineno,amsfonts}
\usepackage{graphicx}   
\usepackage{amsmath}
\usepackage{bbm}
\usepackage{mathrsfs}
\usepackage{mathtools}
\usepackage{physics}

\newcommand{\outpr}[2]{\vert{#1}\rangle\langle{#2}\vert}
\newcommand{\proj}[1]{\vert{#1}\rangle \langle{#1}\vert}

\usepackage{orcidlink}
\newcommand{\commute}[2]{\left[{#1},{#2}\right]}
\newcommand{\expv}[2]{\left\langle{#1}\right\rangle_{#2}}
\newcommand{\dmel}[2]{\left\langle{#1}\left\vert{#2}\right\vert{#1}\right\rangle}

\begin{document}
\title{Experimental prime factorization via the feedback quantum control
} 

\author{K. B. Hari Krishnan}
\email{harikrishnan.kb@students.iiserpune.ac.in}
\affiliation{Department of Physics and NMR Research Center, Indian Institute of Science Education and Research, Pune 411008, India}

\author{Vishal Varma}
\email{vishal.varma@students.iiserpune.ac.in}
\affiliation{Department of Physics and NMR Research Center, Indian Institute of Science Education and Research, Pune 411008, India}

\author{T. S. Mahesh}
\email{mahesh.ts@iiserpune.ac.in}
\affiliation{Department of Physics and NMR Research Center, Indian Institute of Science Education and Research, Pune 411008, India}

\begin{abstract}
Prime factorization on quantum processors is typically implemented either via circuit-based approaches such as Shor’s algorithm or through Hamiltonian optimization methods based on adiabatic, annealing, or variational techniques. While Shor’s algorithm demands high-fidelity quantum gates, Hamiltonian optimization schemes,  with prime factors encoded as degenerate ground states of a problem Hamiltonian, generally require substantial classical post-processing to determine control parameters.
We propose an all-quantum, measurement-based feedback approach that iteratively steers a quantum system toward the target ground state, eliminating the need for classical computation of drive parameters once the problem Hamiltonian is determined and realized. As a proof of principle, we experimentally factor the biprime 551 using a three-qubit NMR quantum register and numerically analyze the robustness of the method against control-field errors. We further demonstrate scalability by numerically implementing the FALQON factorization of larger biprimes, 9,167 and 2,106,287, using 5 and 9 qubits, respectively.
\end{abstract}

\maketitle

\section{Introduction}
Circuit-based implementations of quantum algorithms face significant challenges in the present noisy intermediate-scale quantum (NISQ) era, primarily due to imperfect state preparations and imprecise gate realizations \cite{preskill2018quantum}.
Most notably, Shor's algorithm for prime factorization of an $l$-bit number needs $O(l^3)$ gates with extremely small infidelities of $O(l^{-4})$ \cite{365700,10.5555/2011698.2011703}.
Accordingly, only small-scale demonstrations of Shor's algorithm  with a few qubits have so far  been carried out in various architectures, such as NMR \cite{vandersypen2001experimental}, photonic qubits \cite{lu2007demonstration}, and trapped ions \cite{monz2016realization}.  Therefore, for factoring larger numbers by quantum processors, one resorts to the Hamiltonian optimization method, wherein the composite number is encoded as a problem Hamiltonian $H_p$, whose degenerate ground states represent factors.  Starting from an initial state, the quantum system is gradually driven to the ground state of $H_p$ by quantum control methods
such as adiabatic computing \cite{peng2008quantum,pal2019hybrid}, quantum annealing \cite{jiang2018quantum},  the variational approach \cite{sobhani2025variational}, or its variants like QAOA \cite{anschuetz2019variational,yan2022factoring}.
The quantum optimization approaches bring their own challenges, such as slow evolutions, specific state initialization requirements, or expensive classical computing to design control field parameters that realize time-dependent Hamiltonians. 

Recently, a feedback-based algorithm for quantum optimization (FALQON) was introduced that ensures an asymptotic convergence to the target state \cite{PhysRevLett.129.250502,PhysRevA.106.062414}.  
FALQON is a Lyapunov-inspired iterative quantum algorithm that makes certain measurements of the instantaneous state at every iteration to determine the drive parameters for the subsequent iteration \cite{kuang2008lyapunov,cong2013survey}. 
Starting from almost any initial state, FALQON can iteratively take the system towards the unitarily connected ground state of any physically realizable problem Hamiltonian.
Another key advantage of this method is that measurements of the experimentally realized state at each iteration completely determine subsequent drive parameters, thus effectively eliminating the need for expensive classical computing to design control field profiles. Here, the only requirement is an efficient and robust realization of the time-independent problem Hamiltonian $H_p$.  This one-time task can be efficiently implemented via digital-analog quantum computing (DAQC), which involves single-qubit rotations separated by natural multi-qubit evolutions such as Ising interactions \cite{martin_digital-analog_2020}.  Finally, FALQON iterations self-correct against control errors and thereby are resilient to experimental imperfections.  

Recently, FALQON and its variants \cite{PhysRevResearch.6.033336} have been used for a variety of tasks, such as solving a graph partitioning problem \cite{PhysRevLett.129.250502}, eigenvalue estimation 
using superconducting qubits \cite{rahman2024feedback}, ergotropy estimation by an NMR register \cite{joshi2024maximal}, and realizing an optimized controlled-phase gate in neutral atoms \cite{PhysRevA.109.062603}.  Several theoretical developments have also been made.  These include 
efficient FALQON via shadow measurements
\cite{snht-7jsf},
accelerating FALQON via time rescaling \cite{qc91-5mj2}, imaginary-time-enhanced FALQON for universal ground-state preparation \cite{vanlong2025imaginarytimeenhancedfeedbackbasedquantumalgorithms}.

In this work, we propose utilizing FALQON for the prime factorization task and demonstrate the experimental factoring of a biprime number, 551, using a three-qubit NMR register.  We also analyze the robustness of the FALQON algorithm against experimental errors.
The article is organized as follows. Sec. \ref{sec:Theory} describes constructing the problem Hamiltonian and the Lyapunov theory behind the FALQON algorithm.  Sec. \ref{sec:Expt} describes the experimental factoring of 551 using an NMR register and the robustness analysis. In Sec. \ref{sec:numan}, we numerically analyze the robustness of the feedback method and numerically implement the FALQON factorization of larger biprimes, 9,167 and 2,106,287, using 5 and 9 qubits, respectively.   Finally, Sec. \ref{sec:cncl} summarizes the results and provides an outlook.

\section{Theory \label{sec:Theory}}
\subsection{Constructing the problem Hamiltonian}
Consider an $l_n$ bit biprime number $n$ with prime factors $p$ and $q$ of bit-lengths $l_p$ and $l_q$ respectively, with binary expansion
\begin{eqnarray}
    n =  
    \sum_{i = 0}^{l_{n}-1} 2^{i}n_{i} =pq= \sum_{j,k=0}^{l_p - 1 ,l_q - 1}2^{j+k} p_{j}q_{k}.    \label{eq:BitMulti1}
\end{eqnarray}
The hardest cases are those where the factors are comparable, and for simplicity we assume $l_p=l_q= \lfloor l_n/2\rceil$, where we have rounded $l_n/2$ to the nearest integer.
Excluding nontrivial cases, we can assume that $\{n,p,q\}$ are all odd, with bit 1 as the most- and least-significant binary digits.  Thus, dropping these terminal bits, we only need to determine the remaining $\lfloor l_n/2\rceil-2$ bits. For such a biprime with a pair of five-bit factors, we can construct the bitwise multiplicative table as follows \cite{pal2019hybrid}.

\begin{center}
$
\begin{array}{c|c c c c c c c c c c}
               551     & 1   & 0   & 0   & 0   & 1   & 0   & 0   & 1   & 1   & 1 \\
         n  &  n_9  &  n_8  &  n_7  &    n_6    &    n_5    &    n_4    &    n_3    &    n_2    &  n_1  &  n_0   \\
\hline p  &       &       &       &           &           &   1       &  p_3      &  p_2      &  p_1  &  1     \\
       q  &       &       &       &           &           &   1       &  q_3      &  q_2      &  q_1  &  1     \\
\hline r_0 &       &       &       &           &           &   1       &  p_3      &  p_2      &  p_1  &  1     \\
       r_1 &       &       &       &           &  q_1      &  p_3 q_1  &  p_2 q_1  &  p_1 q_1  &  q_1  &        \\
       r_2 &       &       &       &  q_2      &  p_3 q_2  &  p_2 q_2  &  p_1 q_2  &  q_2      &       &        \\
       r_3 &       &       &  q_3  &  p_3 q_3  &  p_2 q_3  &  p_1 q_3  &  q_3      &           &       &        \\
       r_4 &       &   1   &  p_3  &  p_2      &  p_1      &  1        &           &           &       &        \\ 
\hline \text { carry } & c_9 & c_8 & c_7 & c_6 & c_5 & c_4 & c_3 & c_2 & c_1 & 0 
\end{array}
$
\end{center}
From the above table, one obtains the constraint \cite{pal2019hybrid} 
\begin{eqnarray}
    f_m = \displaystyle\sum_{k = k_d}^{k_u}  p_{m-k}q_{k} + c_{m-1} - 2c_m - n_m = 0,
    \label{eq:CoupledEq2}
\end{eqnarray}
where $k_d = \max(0,m-l_p+1)$ and $k_u=\min(m,l_q-1)$.
Thus, the optimization problem is to minimize the energy,
\begin{eqnarray}
    E \equiv \abs{f_{0}}^{2} + \abs{f_{1}}^{2} +  ...  + \abs{f_{l_{p} + l_{q} - 2}}^{2},    \label{eq:HpGeneral}
\end{eqnarray}
which can be modeled as the ground state of a problem Hamiltonian $H_p$.
For the specific case of $n=551$, one can encode the problem Hamiltonian onto a three-qubit register \cite{pal2019hybrid}
\begin{eqnarray}
    H_p =  \frac{1}{4}\left(\sigma_{z1}\sigma_{z2}  -  \sigma_{z2}\sigma_{z3}  +  \sigma_{z1}\sigma_{z3}\right), 
    \label{eq:Hp}
\end{eqnarray}
whose ground states reveal factors after bit reversal and appending with bit 1 on the most and least significant places.

\subsection{Convex Optimization using FALQON \label{sec:TheoryFAlQON}}
Consider the qubit register evolving with the total Hamiltonian $H_p+\beta(t)H_d$, where $H_d$ is a suitable drive Hamiltonian with a scalar time-dependent control parameter $\beta(t)$.  In the iterative method, instead of the continuous function $\beta(t)$, we use discrete drive parameters $\{\beta_1,\beta_2,\cdots ,\beta_n\}$,  which need to be determined.
Suppose the system is prepared in a state $\rho_j$ with energy $\expv{H_p}{\rho_{j}}$ after iteration $j$.  In the subsequent iteration, our goal is to further reduce the energy $\expv{H_p}{\rho_{j+1}}$ by evolving $\rho_j$ under the problem Hamiltonian $H_p$ and the drive Hamiltonian $\beta_{j+1} H_d$. According to the Liouville von Neumann equation
\begin{equation}
\dot{\rho}_j = -i\commute{H_p+\beta_{j+1}H_d}{\rho_j},
\end{equation}
where we have set $\hbar=1$ for brevity. In order to determine the necessary drive parameter $\beta_{j+1}$ that reduces the energy,
FALQON treats the energy 
\begin{equation}
E_j = \expv{H_p}{\rho_j} =\Tr{H_p {\rho_j}}    
\end{equation}
as the control Lyapunov function \cite{kuang2008lyapunov,cong2013survey} whose time derivative is
\begin{align}
\dot{E}_j &=  \Tr{H_p \dot{\rho_j}} = 
-i\Tr{H_p \commute{H_p+\beta_{j+1} H_d}{\rho_j}}
\nonumber \\
&= -\beta_{j+1} \expv{C}{\rho_j},~\mbox{where,}~
C = 
i\commute{H_p}{H_d}.
\end{align}
In the above, we have used cyclic permutations within the trace operation.  Note that $\expv{C}{\rho_j}$ is real-valued, being the expectation value of the Hermitian observable $C$.
The key idea of FALQON is to measure $\expv{C}{\rho_j}$ at every iteration $j$ and set the $(j+1)$th control as
\begin{align}
\beta_{j+1} = c\,\expv{C}{\rho_j},
\end{align}
with a suitable scaling factor $c$ such that,
\begin{align}
\dot{E}_j = - c\, \expv{C}{\rho_j}^2 \leq 0,
\end{align}
which leads to energy minimization \cite{PhysRevLett.129.250502,PhysRevA.106.062414}. Thus, all we need to do in every iteration is to efficiently implement $H_p$, $H_d$, and then measure $\expv{C}{\rho_j}$, which tells what drive to apply in the next iteration.  This summarizes the idea behind FALQON.
In the following, we describe the experimental factoring of 551 by implementing FALQON on a three-qubit NMR register.

\section{Experimental factoring of 551 \label{sec:Expt}}
\subsection{The experimental setup}
\begin{figure*}
    \includegraphics[trim=0cm 0cm 0cm 0cm,clip=,width=\linewidth]{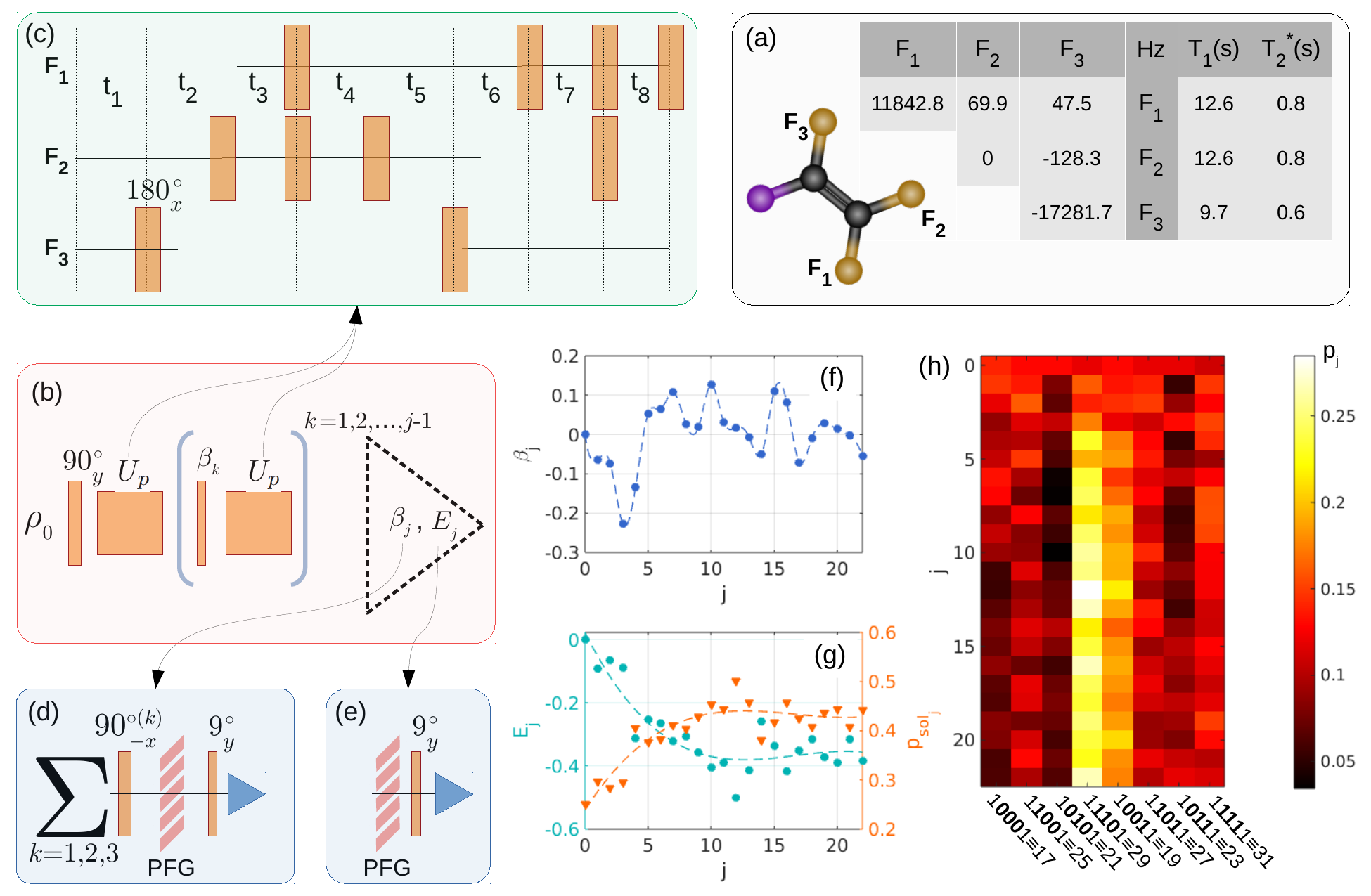}
    \caption{(a) Molecular structure of 1,1,2-trifluoro-2-iodoethene and table of NMR parameters, namely resonance offsets $\nu_i$ (diagonal elements), and scalar couplings $J_{ij}$ (off-diagonal elements), and the relaxation times $T_1$ and $T_2^*$. (b) The FALQON experimental scheme.  (c) DAQC implementation of $\exp(-i~H_p~0.2)$.
    Here, durations $t_1$ to $t_8$ are 1.72, 1.2, 0.82, 1.54, 0.8, 0.88, 0.62, 1.14 ms. (d) Measuring the commutator observable $C$ by a set of three diagonal state tomography experiments and thereby evaluating the drive parameter $\beta_j$. (e) Measuring the energy $E_j$ using the diagonal state tomography.   (f) Experimentally obtained drive parameter $\beta_j$ (in degrees) plotted versus $j$. (g) Experimentally measured energy $E_j$ and the solution-space probability $p_\mathrm{sol}$. (h) Heatmap of probabilities of basis states (bit-reversed and padded with bit-1 in the least and most significant places). Note that the states corresponding to the factors 19 and 29 are clearly more populated than the other states.}
    \label{fig:Results}
\end{figure*}

All the experiments were carried out on a Bruker 500 MHz NMR spectrometer at an ambient temperature of 300 K. Fig. \ref{fig:Results} shows all the important aspects of the experiment.  We use the three spin-1/2 $^{19}$F nuclei of 1,1,2-trifluoro-2-iodoethene  as the three-qubit register.
The molecular structure of the system, resonance offsets $\nu_j$, J-couplings $J_{ij}$, and $T_1$, $T_2^*$ relaxation times are shown in Fig. \ref{fig:Results} (a).  The system Hamiltonian of the register, in the rotating frame of the RF carrier, is of the form \cite{levitt2008spin}
\begin{align}
H_\mathrm{sys} = -2\pi \sum_{i=1,2,3}\nu_i\sigma_{zi}/2+
2\pi \sum_{i<j}J_{ij}\sigma_{zi}\sigma_{zj}/4.
\end{align}

\subsection{The FALQON experimental scheme}
The FALQON iteration protocol is shown in Fig. \ref{fig:Results} (b). Unlike the adiabatic methods, which demand specific initial states to avoid level-crossings during the dynamics, FALQON has more freedom in selecting the initial state. In fact, for the present task of factoring via Hamiltonian optimization, we simply start with the NMR thermal state, 
\begin{equation}
\rho_0 = \frac{1}{2} \sum_{i=1}^3 \sigma_{zi},   
\end{equation}
and convert it into an observable transverse magnetization using a $90^\circ_y$ pulse.  Thus, no elaborate preparation of pseudopure states is required, which is another major advantage over the adiabatic methods.

Following Ref. \cite{PhysRevLett.129.250502}, we apply alternating segments of $H_p$ and $H_d$, which is a Trotter approximation of the simultaneous evolution. Note that a diagonal problem Hamiltonian can be efficiently realized via DAQC algorithm \cite{parra-rodriguez_digital-analog_2020-1}. One possible implementation of $U_p = \exp(-iH_p\delta)$ is shown in Fig. \ref{fig:Results} (c).  The main issue with this solution is its sensitivity to imperfections in the $\pi$ pulses.  Therefore, to achieve high-fidelity  operation in the presence of experimental limitations, we have  utilized the gradient ascent pulse engineering (GRAPE) \cite{khaneja_optimal_2005-1} and generated a robust RF sequence realizing $U_p$ for $\delta=0.2$ with a fidelity of 0.98 with up to 20\% RF inhomogeneity.

While there are several possible options for the drive Hamiltonian, we choose the simplest option, 
\begin{align}
H_d = \frac{1}{2}\sum_{i=1}^3\sigma_{xi}
\label{eq:Hd}
\end{align}
and the corresponding drive unitary for the $k$th iteration of the form
\begin{align}
U_{d,k} = \exp(-i\beta_k H_d),
\end{align}
which is denoted by $\beta_k$ in Fig. \ref{fig:Results}
 (b).
 
The next important step is to efficiently measure the commutator observable $\expv{C}{\rho_j}$.  Of course, the brute force method is to perform the full-state tomography  to determine $\rho_j$ and then evaluate the expectation value \cite{nielsen2010quantum}.  However, in general, the full-state tomography requires an exponentially large number of measurements and hence is not scalable \cite{PhysRevA.87.062317}.  Instead, if we can convert the problem into diagonal tomography, which involves measuring only the probabilities, it can be efficiently implemented in most architectures.  In fact, in NMR, the diagonal tomography involves only one linear measurement using a small flip-angle pulse after a pulsed field gradient (PFG) dephases all the coherences \cite{suter2008spins}. Thus, to efficiently measure $C$, we first evaluate it using Eqs. \ref{eq:Hp} and \ref{eq:Hd} and express it in terms of diagonal operators,
\begin{gather}
C = i\commute{H_p}{H_d}  
= \sum_{k=1,2,3} C_k,~\mbox{where}~
C_k = R_k D_k R_k^\dagger,
\nonumber \\
R_k = \exp(-i\pi/2\, \sigma_{xk}/2),~ \mbox{is  $90^\circ_x$ rotation of $k$th qubit},
\nonumber\\
\mbox{with}~D_1 = \left(\sigma_{z1}\sigma_{z2}+\sigma_{z1}\sigma_{z3}\right)/2,
\nonumber \\
~~~~~ D_2 = \left(\sigma_{z1}\sigma_{z2}-\sigma_{z2}\sigma_{z3}\right)/2,
~\mbox{and}~
\nonumber \\
~~~~~ D_3 = \left(-\sigma_{z2}\sigma_{z3}+\sigma_{z1}\sigma_{z3}\right)/2.
\end{gather}
Note that $R_k$ are chosen such that $D_k = \sum_j d_j^{(k)} \proj{j}$ are diagonal operators.  The individual expectation values are
\begin{align}
\expv{C_k}{\rho} &= 
\Tr{R_k D_k R_k^\dagger \, \rho}
\nonumber \\
&= \Tr{D_k \, {\rho}^{(k)}},~
\mbox{where}~{\rho^{(k)}} = R_k^\dagger \rho R_k
\nonumber \\
&=\Tr{\sum_j d_j^{(k)} \proj{j} \, \sum_{lm} {\rho^{(k)}}_{lm} \outpr{l}{m} } 
\nonumber \\
&=\Tr{ \sum_{jm} d_j^{(k)} {\rho^{(k)}}_{jm}  \,  \outpr{j}{m} } = \sum_{m} d_m^{(k)} {\rho^{(k)}}_{mm}.
\end{align}
Thus, the diagonal tomography of ${\rho^{(k)}}$ enables us to measure $\expv{C_k}{\rho}$, and carrying out three such measurements yields $\expv{C}{\rho}$.  The pulse sequence for this part of the experiment is shown in Fig. \ref{fig:Results} (d).

Finally, it is useful to monitor the energy 
of the state $\rho$ w.r.t. the Hamiltonian 
$H_p = \sum_m e_m \proj{m}$.  The energy is expressed as
\begin{align}
E(\rho) &= \Tr{H_p \rho} = 
\Tr{\sum_m e_m \proj{m} \sum_{kl}\rho_{kl}\outpr{k}{l}}
\nonumber \\
&=\Tr{\sum_{ml} e_m \rho_{ml} \outpr{m}{l}}
=\sum_{m} e_m \rho_{mm}.
\end{align}
Here, a single diagonal tomography suffices, and no $R_k^\dagger$ rotation is needed.  The corresponding pulse sequence is shown in Fig. \ref{fig:Results} (e).  In the following, we describe experimental results.
 
\subsection{Experimental results \label{sec:Results}}
Fig. \ref{fig:Results} (f) plots the experimentally obtained drive parameter $\beta_j$ (in degrees) versus the iteration number $j$.  Here, $\beta_0$ was set to $0$, and subsequent $\beta_j$ values were obtained via the FALQON scheme described above with the scaling factor set to $c=0.25$.  After an initial large oscillation, $\beta_j$ values appear to approach values closer to 0, as expected.  Fig. \ref{fig:Results} (g) plots the energy $E_j$ (filled circles; in arbitrary scale) versus the iteration $j$.  Although there are fluctuations, it is clear that the energy is following a decreasing trend and then saturating.  An overlaying plot shows the probabilities
\begin{align}
p_\mathrm{sol_j} &= \dmel{011}{\rho_j} +\dmel{100}{\rho_j}
\end{align}
of the solution space.
We see that the $p_\mathrm{sol_j}$ starts from about 0.2 and gradually settles at about 0.4.  Finally, Fig. \ref{fig:Results} (h) shows the heatmap of probabilities $p_m$ of various basis states (bit-reversed and padded with bit-1 in the least and most significant places), plotted against the iteration number $j$. We can observe a uniform initial probability distribution concentrating mainly on two states $\ket{011}$ and $\ket{100}$.  After bit-reversal and padding bit 1 on the least and the most significant places, the resulting binary strings represent the prime factors of 551, namely $10011 \equiv 19$ and $11101 \equiv 29$.
It is convincing to note that FALQON maintains higher probabilities on these states with a clear contrast over other states for the last 16 iterations.  This completes the experimental demonstration of factoring by the feedback method.  In the following we describe the robustness of the method over different types of errors.

\section{Numerical analysis \label{sec:numan}}
\subsection{Robustness}
\begin{figure}
    \includegraphics[trim=0cm 0cm 0cm 0cm,clip=,width=1\linewidth]{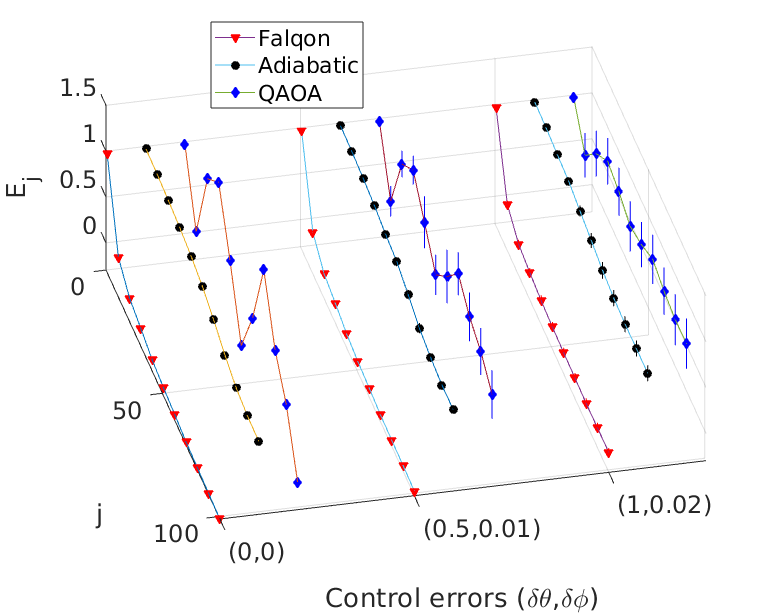}
    \caption{Energy $E_j$ plotted versus the iteration number $j$ and control errors $(\delta \theta, \delta\phi)$ for three optimization algorithms as indicated. 
    \label{fig:noiserobustness}}
\end{figure}

We now numerically analyze the robustness of  FALQON factorization against random noises, control-field inhomogeneities, and nonoptimal step sizes.  Fig. \ref{fig:noiserobustness} compares the energy $E_j = \expv{H_p}{\rho_j}$ with iteration number $j$ and control errors $(\delta \theta, \delta\phi)$ for the factoring task employing one of the three optimization algorithms, namely FALQON, the adiabatic method \cite{farhi2014quantum}, and QAOA \cite{hadfield2019quantum}.  Here $\delta \theta$ and $\delta \phi$ represent the amplitude of the random errors in all the pulse angles $\theta$ and pulse phases $\phi$. It is evident that FALQON achieves faster convergence than the adiabatic method.  The QAOA was not monotonic, at least for the parameters used in our simulations.  The robustness of FALQON against the random noise is not surprising, since every past error is naturally accounted for while determining the subsequent drive parameters.  

\begin{figure}
    \includegraphics[trim=0cm 0cm 0cm 0cm,clip=,width=1\linewidth]{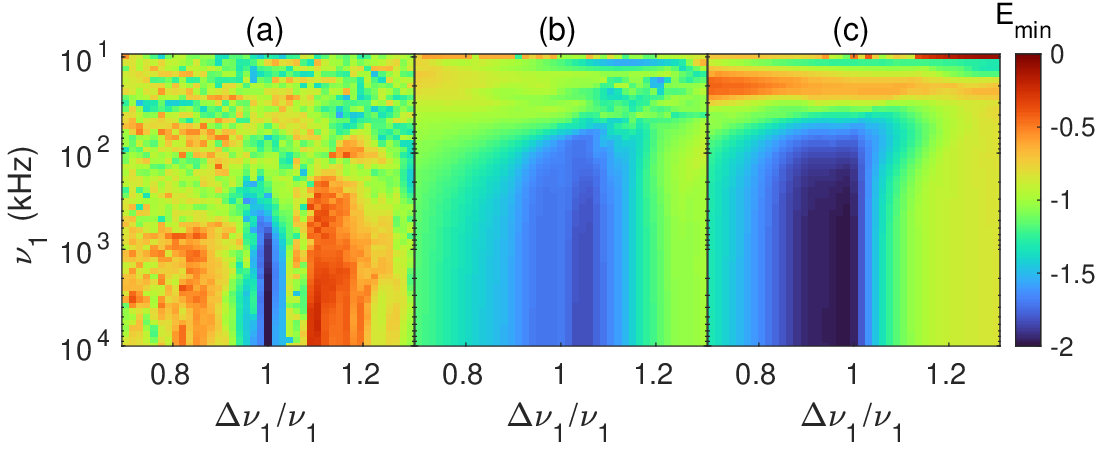}
    \caption{Colormap of the minimum energy reached over 100 FALQON iterations versus RF amplitude $\nu_1$ and the RF inhomogeneity $\Delta \nu_1/\nu_1$ using (a) $H_p$ realized by DAQC, (b) $H_p$ realized by GRAPE, and (c) theoretical $H_p$. }
\label{fig:rfirobustness}
\end{figure}

In NMR, the limited control field amplitude and the inhomogeneity in the control field over the volume of the NMR sample are more serious than random errors \cite{cavanagh1996protein}.  For instance, these two limitations render the DAQC $\pi$ pulses imprecise.  Fig. \ref{fig:rfirobustness} plots colormaps for minimum energy reached over 100 FALQON iterations against the maximum RF amplitude ($\nu_1$) and RF inhomogeneity (RFI) $\Delta \nu_1/\nu_1$, for factoring under three different scenarios, namely Fig. \ref{fig:rfirobustness} (a) $H_p$ realized by DAQC,  Fig. \ref{fig:rfirobustness} (b) $H_p$ realized by GRAPE, and Fig. \ref{fig:rfirobustness} (c) theoretical $H_p$.  In all the cases, we use realistic drive pulses.
It is clear that DAQC realization of $H_p$ is computationally scalable, but it needs a large number of close-to-ideal $\pi$ pulses to succeed.  However, the GRAPE realization of $H_p$ is designed to work with limited RF amplitude and to be robust against RFI.  Accordingly, Fig. \ref{fig:rfirobustness} (b) corresponding to factoring with the GRAPE sequence, shows a wider operational area, almost like the ideal $H_p$.

\begin{figure}
    \includegraphics[width=1\linewidth]{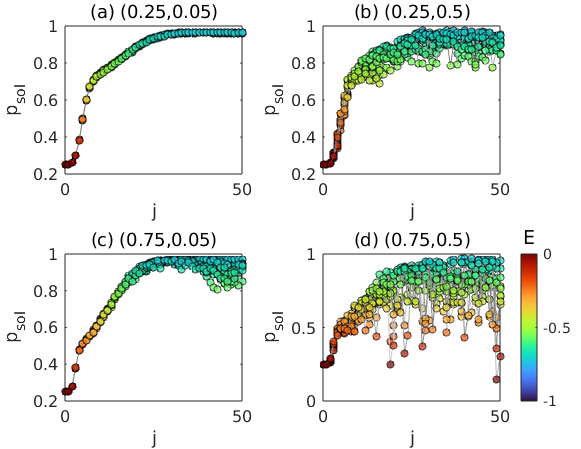}
    \caption{Ten independent trajectories of probability of solutions $p_\mathrm{sol}$ versus FALQON iteration number $j$ for different parameters $(c,\delta \theta)$ controlling the step size and control-field noise. The energy $E_j$ is indicated by the filling colors of the circles.}
    \label{fig:solvsnonsol}
\end{figure}
Finally, it is important to understand the step size and how it affects the robustness.  Fig. \ref{fig:solvsnonsol} overlays 10 independent trajectories of the probabilities 
$p_\mathrm{sol}$
of the solution space versus the FALQON iteration number for different parameters
$(c,\delta\theta)$, where $c$ is the step size and $\delta\theta$ is the amplitude of the random noise in the flip angles due to the control field noise. The filling colors of circles indicate the energy $E_j = \expv{H_p}{\rho_j}$ of the state.  Note that with a smaller step size $c=0.25$, FALQON is robust against the random noise.  However, with increased step size $c=0.75$, it has become sensitive to the random noise.  Note the similarity of our experimental results in Fig. \ref{fig:Results} (g) with Fig. \ref{fig:solvsnonsol} (b).  Thus, while too small step size may make FALQON inefficient, too large step size will make it less robust.  This completes the robustness analysis of the FALQON-based prime factorization.  
 
\subsection{Factoring larger numbers}

\begin{figure*}
     \includegraphics[width=1\linewidth]{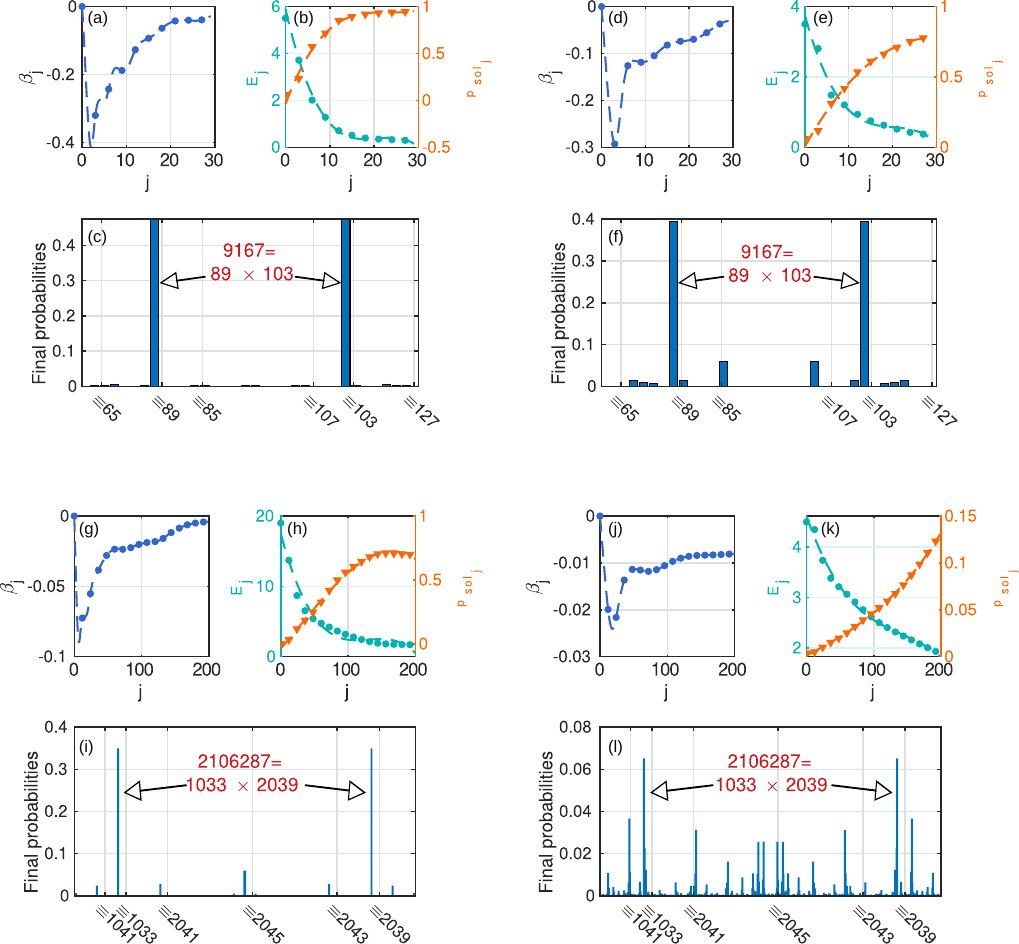}
     \caption{(a-f) Numerical The progression of drive parameter $\beta_j$, energy $E_j$, solution-space probability $p_{\mathrm{sol}_j}$, and final probabilities for the FALQON factoring of 9,167 with full (a-c) and truncated (d-f) Hamiltonians.  (g-l) Similar progressions and final probabilities for the FALQON factoring of 2,106,287 with full  (g-i) and truncated (j-l) Hamiltonians.  }
    \label{fig:FALQON_FactLargerNo}
\end{figure*}

We now numerically implement the FALQON factorization of larger biprimes, namely 9,167 and 2,106,287, with 5- and 9-qubits, respectively. In both cases, we choose the initial state $\rho_0 = \ket{0}^{\otimes n}$ and the drive Hamiltonian $H_d= \frac{1}{2}\sum_{i=1}^n \sigma_{xi}$, where $n$ is the number of qubits.  In the following, we denote $\sigma_{zk}$ by $z_k$ to save the symbols.  For 9,167, the full problem Hamiltonian is,
\begin{gather}
H_p^{(9167)} = 
   {z_1} (- {z_2}
+   {z_3}
+ 2  {z_4})
+ 2 {z_2} ( {z_3}
-   {z_5}) 
-  {z_3} ( 2 {z_4}
-     {z_5})
\nonumber \\
+   {z_4}  {z_5}
+   {z_1}  {z_2} ( {z_3}  {z_4}
+    {z_4}  {z_5})
+   {z_2}  {z_3}  {z_4}  {z_5}.
\label{eq:hp9167}
\end{gather}
Although the above problem Hamiltonian consists of many-body terms, they can be realized via standard techniques, such as the algorithmic approach of Ref.
\cite{ajoy2012algorithmic}.
 Results of numerical simulations are shown in Fig. \ref{fig:FALQON_FactLargerNo}. Figs. \ref{fig:FALQON_FactLargerNo} (a-c) corresponding to the biprime 9,167, show the progression of drive parameter $\beta_j$, energy $E_j$, solution-space probability $p_{\mathrm{sol}_j}$, and final probabilities of basis states clearly concentrated at prime factors 89 and 103.  
 
 Now we may ask if all the terms of $H_p^{(9167)}$ in Eq. \ref{eq:hp9167} are crucial for factoring.  To this end, we found that the truncated Hamiltonian,
\begin{gather}
H_{p ~\mathrm{(trunc)}}^{(9167)} =
   {z_1} (- {z_2}
+ 2  {z_4})
+ 2 {z_2} ( {z_3}
-   {z_5}) ,
\end{gather}
which retains only four terms of $H_p^{(9167)}$, also has the same degenerate ground states that correspond to the factors of 9,167. Moreover, FALQON is able to capture the factors with the same number of iterations as shown in Figs. \ref{fig:FALQON_FactLargerNo} (d-f).
Thus, if one can find a systematic way of truncating the problem Hamiltonian, it will further simplify the practical implementation of factoring.

Now for 2,106,287, the full problem Hamiltonian is

\begin{gather}
H_p^{(2106287)} = 
   {z_1}(- {z_2}
+    {z_3}
- 2  {z_5}
-    {z_6}
-    {z_7}
- 2  {z_8}
- 2  {z_9})
\nonumber \\
+ {z_2}(- 2 {z_4}
-    {z_5}
-    {z_6}
- 2  {z_7}
- 2  {z_8}
- 2  {z_9})
+   {z_3}(- {z_4}
-    {z_5}
\nonumber \\
- 2  {z_6}
- 2  {z_7}
- 2  {z_8}
-    {z_9})
+ {z_4}(- 2 {z_5}
- 2  {z_6}
- 2  {z_7}
-    {z_8}
- 3  {z_9})
\nonumber \\
+ {z_5}(- 2 {z_6}
-    {z_7}
- 3  {z_8}
- 2  {z_9})
+ {z_6} (- 3 {z_7}
- 2  {z_8}
- 2  {z_9})
\nonumber \\
+ {z_7}(- 2 {z_8}
-    {z_9})
-   {z_8} {z_9}
+   {z_1} {z_2} ({z_3} {z_4}
+    {z_4} {z_5}
+    {z_5} {z_6}
+    {z_6} {z_7}
\nonumber \\
+    {z_7} {z_8}
+    {z_8} {z_9})
+   {z_1} {z_3} ({z_4} {z_6}
+    {z_5} {z_7}
+    {z_6} {z_8}
+    {z_7} {z_9})
\nonumber \\
+   {z_1} {z_4} ( {z_5} {z_8}
+   {z_6} {z_9})
+   {z_2} {z_3}( {z_4} {z_5}
+    {z_5} {z_6}
+    {z_6} {z_7}
+    {z_7} {z_8}
\nonumber \\
+    {z_8} {z_9})
+   {z_2} {z_4} ({z_5} {z_7}
+   {z_6} {z_8}
+   {z_7} {z_9})
+   {z_2} {z_5} {z_6} {z_9}
+   {z_3} {z_4} ( {z_5} {z_6}
\nonumber \\
+   {z_6} {z_7}
+   {z_7} {z_8}
+    {z_8} {z_9})
+   {z_3} {z_5} ({z_6} {z_8}
+   {z_7} {z_9})
+   {z_4} {z_5} ({z_6} {z_7}
\nonumber \\
+   {z_7} {z_8}
+   {z_8} {z_9})
+   {z_4} {z_6} {z_7} {z_9}
+   {z_5} {z_6} ({z_7} {z_8}
+    {z_8} {z_9})
+   {z_6} {z_7} {z_8} {z_9}.
\end{gather}
Figs. \ref{fig:FALQON_FactLargerNo} (g-i) show the progression of $\beta_j$, $E_j$,  $p_{\mathrm{sol}_j}$, and the final probabilities of basis states clearly concentrated at prime factors 1033 and 2039.  
Here again, the truncated Hamiltonian 
\begin{gather}
H_{p ~\mathrm{(trunc)}}^{(2106287)} = 
-2 {z_2}{z_7}
+   {z_2} {z_3}({z_1}  {z_4}+ {z_4} {z_5}
+    {z_5} {z_6}
+    {z_6} {z_7}
\nonumber \\
+    {z_7} {z_8}
+    {z_8} {z_9})
+   {z_3} {z_4}  {z_5} {z_6},
\end{gather}
has the same degenerate ground states as that of the full Hamiltonian, and therefore FALQON 
can determine the factors with high probability, as shown in Figs. \ref{fig:FALQON_FactLargerNo}
(j-l).

 Although here we have used a single-drive Hamiltonian, one may employ multiple-drive Hamiltonians that may improve convergence rate. Also, since verifying factors is an easy task, one may stop at far fewer iterations than shown in Fig. \ref{fig:FALQON_FactLargerNo}.

\section{Conclusions \label{sec:cncl}}
In this work, we have reported quantum prime factorization using the feedback method known as FALQON.  It is an iterative method of Hamiltonian optimization, where each iteration applies the problem Hamiltonian and a drive Hamiltonian, followed by a measurement that determines the drive parameters for the next iteration. As long as we can efficiently implement the problem and drive Hamiltonians, FALQON involves no classical computation to determine the quantum dynamics.  In this sense, it is an all-quantum algorithm.

Using the FALQON algorithm, we have experimentally demonstrated factoring the biprime number 551 on a three-qubit NMR quantum register.  It is worth noting that, unlike adiabatic methods, which require specific initial states to avoid level crossings, FALQON is not very specific to initial states.  In fact, in our experiments, we directly started from the thermal state of the NMR register.  We explained efficiently realizing the problem Hamiltonian via the DAQC algorithm and its robust implementation using the GRAPE technique.
Although there was sufficient contrast to determine the factors by the 4th  FALQON iteration itself, we continued up to 22  iterations to convincingly observe the lower saturation of energy and upper saturation of the solution-space probability. 

We carried out detailed numerical analysis to understand the effects on FALQON factorization by control field noise, control field inhomogeneities, as well as the step size. We observed FALQON to show faster convergence than the adiabatic method and lesser oscillations than QAOA under different noise amplitudes of the control field.  We have shown that the DAQC realization of the problem Hamiltonian is sensitive to the imperfections of the $\pi$ pulses, while the GRAPE implementation is highly robust.  We have also shown that if the FALQON step sizes are large, it becomes less robust against the control-field noises, and therefore it is important to select a good range for the step size.  Finally, we have demonstrated the scalability of FALQON factorization by numerically factoring larger biprimes, 9,167 and 2,106,287, using 5 and 9 qubits, respectively. Moreover, it is promising to note that FALQON with truncated problem Hamiltonians can also be used to extract factors of large numbers. A systematic method for truncating the problem  Hamiltonians remains an open problem.

As an outlook, we envision the FALQON algorithm for a variety of other quantum information tasks in various architectures. It may also be possible to realize a hybridization of FALQON with other quantum control approaches.

 \section*{Acknowledgements}
Authors acknowledge valuable discussions with Dr. Priya Batra and Dr. Jitendra Joshi.  Valuable suggestions from Dr. Gluza Marek Ludwik of NUS, Singapore, are gratefully acknowledged. H. K.   acknowledges 
the Council of Scientific and Industrial Research (CSIR) fellowship 09/0936(15709)/2022-EMR-I, and V. V. acknowledges the University Grants Commission (UGC) fellowship MR22031373.   We thank the National Mission on Interdisciplinary Cyber-Physical Systems for funding from the DST, Government of India, through the I-HUB Quantum Technology Foundation, IISER-Pune. 

\bibliography{MyLibraryWIthoutBOM}

\end{document}